# The Effect of Particle Size and Concentration on Low-Frequency Terahertz Scattering in Granular Compacts

Keir N. Murphy [1,2], Mira Naftaly [3], Alison Nordon [2,4] and Daniel Markl [1,2]

[1] Strathclyde Institute of Pharmacy & Biomedical Sciences, University of Strathclyde, Glasgow G1 1XQ, UK;
[2] Centre for Continuous Manufacturing and Advanced Crystallisation (CMAC), University of Strathclyde, Glasgow G1 1RD, UK;
[3] National Physical Laboratory, Teddington TW11 0LW, UK;
[4] WestCHEM, Department of Pure and Applied Chemistry, University of Strathclyde, Glasgow G1 1XL, UK

## Abstract

Fundamental knowledge of scattering in granular compacts is essential to ensure accuracy of spectroscopic measurements and determine material characteristics such as size and shape of scattering objects. Terahertz time-domain spectroscopy (THz-TDS) was employed to investigate the effect of particle size and concentration on scattering in specially fabricated compacts consisting of borosilicate microspheres in a polytetrafluoroethylene (PTFE) matrix. As expected, increasing particle size leads to an increase in overall scattering contribution. At low concentrations, the scattering contribution increases linearly with concentration. Scattering increases linearly at low concentrations, saturates at higher concentrations with a maximum level depending on particle size, and that the onset of saturation is independent of particle size. The effective refractive index becomes sublinear at high particle concentrations and exceeds the linear model at maximum density, which can cause errors in calculations based on it, such as porosity. The observed phenomena are attributed to the change in the fraction of photons propagating ballistically versus being scattered. At low concentrations, photons travel predominately ballistically through the PTFE matrix. At high concentrations, the photons again propagate ballistically through adjacent glass microspheres. In the intermediate regime, photons are predominately scattered.

## Introduction

Granular compacts are employed in various industries [1]–[4], including but not limited to the pharmaceutical, food and chemical industries. These industries utilize granular compacts in multiple forms, from drug delivery vectors to producing catalytic materials and breakfast cereals [4], [5]. Analytical techniques have been applied to control the quality and precision of such products, primarily with the aid of spectroscopic techniques. With spectroscopic measurements, various chemical and physical material properties can be determined, such as chemical structure, material homogeneity, and the inclusion of impurities or air voids introduced by the manufacturing process [6]–[9]. The accuracy and precision of spectroscopic measurements can be significantly reduced by measurement errors caused by light scattering, which may lead to inaccurate interpretations and conclusions. The scattering of light in granular compacts is complex [10]–[12]; rather than address it, data analysis often assumes little or negligible scattering, even when it is present and affects the accuracy of measurements. This study experimentally investigates light scattering in granular compacts to elucidate the effect of particle size and concentration on extracted scattering contributions.

Scattering arises when light travels through an inhomogeneous medium with a varying refractive index[13]–[15], and results in transmission loss. The magnitude of scattering loss depends on the wavelength of light relative to the distance of the scattering centers and on the differences in the refractive index. The ratio between wavelength and the size of scattering particles determines scattering behavior and the choice of model that is applied to the quantification of scattering. In cases where the distance between the scattering centers are comparable to the wavelength of the propagating light, Mie theory is applicable[16].

Terahertz time-domain spectroscopy (THz-TDS) has been deployed frequently in the characterization of granular compacts due to its unique ability to determine both the refractive index (RI) and loss coefficient [3], [17], [18] simultaneously. The loss coefficient ($\alpha_{loss}$) is the sum of absorption ($\alpha_{abs}$) and scattering contributions ($\alpha_{scatt}$):

$$\alpha_{loss} = \alpha_{abs} + \alpha_{scatt} \quad (1)$$

Therefore, if the absorption of the solid media is known, then the scattering component can be obtained by simple subtraction.

Scattering in THz-TDS measurements has been previously investigated by both empirical fitting [19] and through a first principles scattering model originally derived by Raman [20]–[22]. Shen et al. modelled the loss spectra ($\alpha_{loss}(v)$) of granular compacts containing polyethylene powder and found that the spectra could be best modelled as a function of frequency ($v$) using [19]

$$\alpha_{loss}(v) = Bv^A \quad (2)$$

with *A* and *B* as fitting parameters. This model has been utilized in various studies, including the non-destructive determination of tablet fragmentation during compaction [23]. Pederson et al. observed the fragmentation of various materials using THz-TDS combined with the model proposed by Shen et al. [19], [23]. This approach, however, did not consider the effect of particle concentration and modelled the total loss coefficient (extinction) rather than the scattering contributions. Franz et al. showed the application of a previously derived first principles model for determining scattering contributions to the loss coefficient. First proposed by Raman to describe the chromatic effects observed by Christiansen, it led to the development of the Christiansen model [12], [14]. This model required significant material knowledge, including the refractive index of both the suspension medium and scattering particle. Using these properties, the model could be fitted using two unknown parameters: the wavefront coverage by the scattering objects and the layer thickness [20], [22]. Once fitted, Franz et al. could show the quantitative value of the estimated scattering contributions to the loss coefficient.

Previous investigations into the effect of an inhomogeneous distribution of particles in paint emulsions have been of significant interest in the past few decades. Fitzwater et al. derived a theoretical model describing light scattering based on Mie theory. They attempted to account for the concentration of inhomogeneities [24]–[26] by assuming a spherical area around the scattering particle dependent on the difference in the refractive index between the matrix and emulsion particles as well as on the wavelength [25]. As particle concentration increases, the interparticle distance decreases; at a concentration level dependent on particle size and wavelength of light, the scattering contributions will saturate.

Latimer et al. [27] employed Mie theory to model the transmission loss in granular compacts and investigate its relationship with particle size and concentration. The study revealed that scattering does not increase linearly with concentration; the scattering particles are closer to each other at higher concentrations, causing photons to travel ballistically through the scattering medium, resulting in a reduction of scattering efficiency. This behavior leads to saturation and a subsequent fall in scattering. Furthermore, they suggested that the particle size is not expected to affect the scattering strength, as the total volume of scattering material remains constant. Latimer et al. also analyzed experimentally cells containing known concentrations of latex spheres suspended in water, showing that the extinction coefficient increased before reaching saturation. However, they could not extract the scattering loss due to a lack of knowledge of the material absorption.

A study by Grinchuk et al. [28] used diffuse reflectance measurements to determine scattering from a painted polyvinyl chloride board. The analysis showed saturation of scattering at higher concentration levels. They applied a percolation theory approach to explain this saturation as a function of the interparticle distance, clarifying that the original methods for the determination of scattering were developed with the scattering of a single particle in mind [28]. This model and experimental evidence of such phenomena have not been investigated for compacted materials using THz-TDS.

This study systematically investigated the effect of particle size and concentration on THz scattering in granular compacts. A set of compacts was devised and fabricated, making it possible to vary the size and concentration of particles in a well-controlled manner. Granular compacts were fabricated using glass microspheres suspended in a PTFE matrix and compacted to create compacts with negligible porosity. Transmission THz-TDS was used to measure frequency-dependent scattering. The study quantified the dependence of the scattering contribution on particle size and concentration. The work studied the power-law dependence of scattering on frequency, identified the dependence of the power-law coefficients on particle size and concentration, and investigated saturation of scattering at higher concentrations.

## Materials & Methods

### *Materials*

Microspheres of borosilicate glass having five size distributions (Cospheric LLC, 2.2 g cm$^{-3}$, 38 – 45, 90 – 106, 125 – 150, 150 – 180 and 180 – 212 µm) were used with PTFE powder (Sigma Aldrich, free flowing, mean particle size 1 µm, 2.2 g cm$^{-3}$) as the matrix material. PTFE was chosen as the matrix material due to its ability to create samples of negligible porosity under sufficient compaction pressure [20], ensuring precise control of the scattering properties of the sample set. The glass spheres were chosen due to their well-controlled particle size distributions, spherical shape, and sufficiently different refractive index from the PTFE matrix material [20].

### *Sample Preparation*

Compacts were fabricated from borosilicate glass microspheres suspended in a PTFE powder matrix using a compaction simulator (HB50, Huxley-Bertram Engineering, Cambridge, UK; London, UK). For each size distribution of the glass microspheres, compacts were produced containing microspheres at six different concentrations in PTFE (2.5, 5, 10, 15, 20 & 30% w/w), and were allowed one week to undergo elastic relaxation to ensure reproducible analysis. The volume fraction, $C$, was subsequently calculated (2, 4, 9, 12, 17, 26 % v/v), using the mass ($m_{glass}$ & $m_{PTFE}$) and true density ($\varrho_{glass}$ & $\varrho_{PTFE}$) of components (Equation 3).

$$C = \frac{m_{glass}/\rho_{glass}}{m_{glass}/\rho_{glass} + m_{PTFE}/\rho_{PTFE}} \qquad (3)$$

Batches of ten plane parallel compacts with a diameter of 9 mm and a thickness of 1 mm were fabricated for each particle size and concentration using 392 MPa of pressure [29]. An additional batch of compacts was fabricated using pure PTFE to obtain the reference optical parameters [20].

## THz-TDS Measurements

THz-TDS measurements were carried out on a commercial system (TeraPulse Lx, Teraview) with a frequency ($v$) resolution of 0.04 THz. All measurements were performed in a nitrogen-purged chamber. Sample thickness ($L$) was measured using a micrometre (±0.005 mm). The frequency-domain field amplitude ($E(v)$) and phase ($\varphi(v)$) for the sample ($s$) and reference ($r$) were obtained from the time-domain data by applying zero padding (next power of 2), apodisation (Hamming approximation) and fast Fourier transform. The frequency-dependent refractive index ($n(v)$) and loss coefficient ($\alpha_{\text{loss}}(v)$) of each sample were calculated using the standard equations [4], [5], [30]–[34].:

$$n(v) = \frac{(\phi_s(v) - \phi_r(v))c}{2\pi v L} \quad (4)$$

$$\alpha_{\text{loss}}(v) = -\frac{2}{L} \ln\left[\frac{(n+1)^2}{4n} \frac{E_s(v)}{E_r(v)}\right] \quad (5)$$

## Extraction of THz Scattering

By combining the known absorption of solid borosilicate glass ($\alpha_{\text{abs,glass}}$) and PTFE ($\alpha_{\text{abs,PTFE}}$), it is possible to estimate the absorption contributions for the compacts at a given concentration ($C$ (%v/v)) of microspheres.

$$\alpha_{\text{abs}} = C\, \alpha_{\text{abs,glass}} + (1 - C)\, \alpha_{\text{abs,PTFE}} \quad (6)$$

Subtracting the estimated absorption contributions from the measured loss coefficient of each compact yields the scattering contribution of the microspheres:

$$\alpha_{\text{scatt}} = \alpha_{\text{loss}}(v) - \alpha_{\text{abs}} \quad (7)$$

## Estimation of THz Scattering at High Concentrations

The optical properties of media with high concentrations of borosilicate microspheres were measured using index-matching fluid. As it was not possible to prepare compacts with a concentration > 30 %v/v of microspheres, a different method was used for these measurements. First, a cuvette was filled with microspheres, their transmission was measured, and the effective refractive index and loss coefficient were calculated. Then the cuvette was filled with paraffin oil ($n_{\text{paraffin}} \sim 1.5$), which has a similar refractive index to PTFE ($n_{\text{PTFE}} \sim 1.45$), allowing for comparison between sample preparation methods. Using the linear mixing formula, the difference between the two values of the effective refractive index provides the volume fraction, $f_{\text{open}}$, of spheres in the cuvette [35].

$$f_{\text{open}} = \frac{\Delta n_{\text{eff}}}{(n_{\text{paraffin}} - 1)} \quad (8)$$

where $\Delta n_{\text{eff}}$ is the refractive index difference between the spheres in air and paraffin.

## Results & Discussion

An example of the frequency-dependent refractive index and loss coefficient of compacts consisting of borosilicate glass microspheres suspended in a PTFE matrix can be seen in Figure 1. The frequency range of valid data was 1.2 THz, the maximum obtainable for these samples given the system's dynamic range [36]. The scattering loss as a function of microsphere concentration for all sphere sizes is plotted at 0.6, 0.8 and 1 THz in Figure 1.

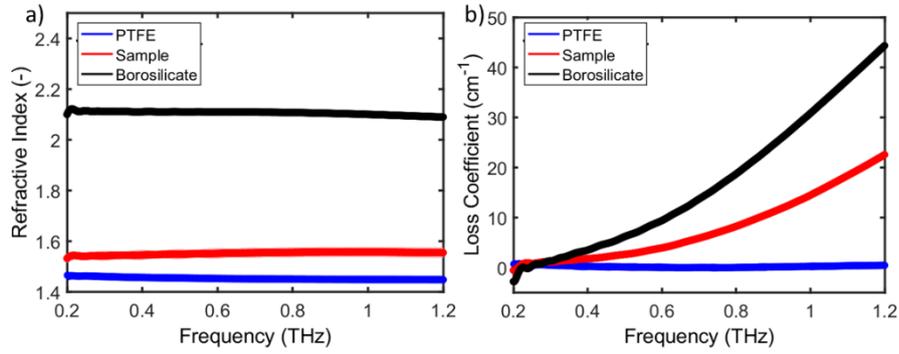

Figure 1: a) Refractive index and b) loss coefficient of compacts consisting of borosilicate glass microspheres (15 %v/v) in a PTFE matrix, pure PTFE samples and borosilicate windows. The values plotted are the mean and standard deviation (number of samples: $N_{\text{sam}} = 10$, $N_{\text{PTFE}} = 10$ & $N_{\text{boro}} = 5$).

The scattering contribution to the measured loss coefficient was evaluated using Eq. 7, and knowledge of the microsphere concentration and the absorption by glass and PTFE. The calculated scattering values are plotted in Figure 2 as a function of concentration for all sphere sizes at three frequencies. It is evident that at low concentrations scattering increases linearly; however, at higher concentrations, scattering becomes saturated. This is attributed to each additional microsphere acting as an individual scattering center at low concentrations, giving rise to a linear dependence of scattering with concentration. At higher concentrations, however, photons will undergo multiple scattering from many microspheres, so that individual microspheres contribute less to the total scattering loss, thus giving rise to saturation.

Further, it is informative to consider photon propagation through the compact, which may occur either ballistically (no scattering) or via scattering interactions. At low concentrations, photons propagate mainly ballistically through the matrix. At high concentrations close to fully dense packing, the photons again propagate ballistically through adjacent glass microspheres. In the intermediate regime, photons are predominately scattered.

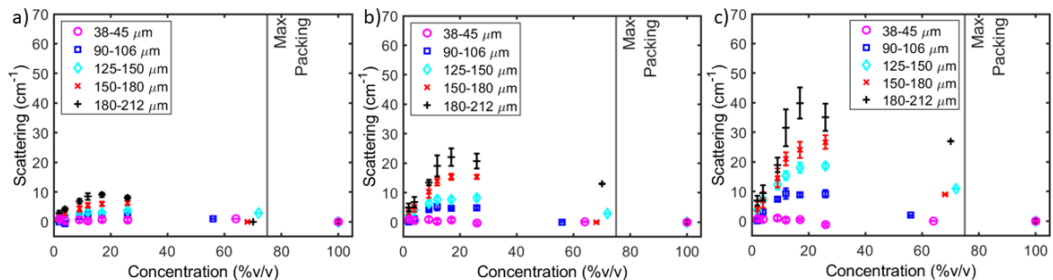

Figure 2: Scattering contributions for each particle size at a) 0.6 THz, b) 0.8THz & c) 1 THz. Values plotted are the mean with error bars representing the standard deviation ($n = 10, < 30\% v/v$). Additionally Maximum packing of spheres (face-centered cubic lattice, 74% v/v) was plotted as a vertical line[37].

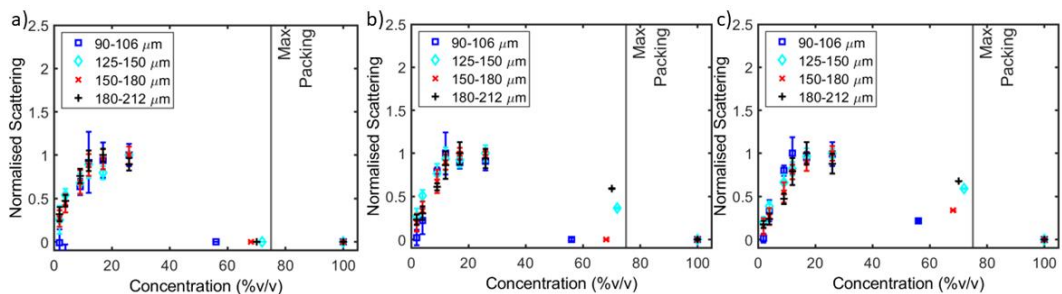

Figure 3: Scattering contributions normalized to the highest value per particle size for a) 0.6 THz, b) 0.8THz & c) 1 THz. Values plotted are the mean with error bars representing the standard deviation ($n = 10, < 30\% v/v$). For clarity, the lowest particle size distribution was removed as the normalization process introduced significant error bars due to the limit of detection. The figure with the lowest particle size is given in Appendix A1. Additionally Maximum packing of spheres (face-centered cubic lattice, 74% v/v) was plotted as a vertical line [37].

As discussed above, scattering is expected to increase linearly at low concentrations until the onset of scattering saturation. When the concentration approaches fully dense packing, scattering is expected to decrease linearly. To experimentally observe the fall of the scattering after the onset of saturation, the method described in Section 2.5 was used to measure scattering for each particle size. This facilitated the evaluation of scattering for concentration levels > 30 % v/v. The normalized curves show that the onset of saturation is independent of both particle size and frequency (Figure 3). In contrast, the drop in scattering at high concentration is size-dependent.

The power law fit followed the methods proposed by Shen et al. with two fitting coefficients (*A & B*, Eq. 2.) However, unlike Shen et al., the power law was applied to the scattering component only and not to the total of scattering and absorption (Figure 4).

Investigating the coefficients obtained from the power law fitting reveals their dependence on particle size and concentration (Figure 5). The pre-exponential factor *B* increases both with concentration and particle size (Figure 5b), reflecting the increase in scattering strength. Exponent *A*, in contrast, increases with particle concentration, but is independent of the size of the spheres (Figure 5a). This agrees with the findings in Figure 3 and shows that the form of scattering behavior is independent of particle size. Once again, we observe a linear increase at low concentrations before the onset of saturation occurs.

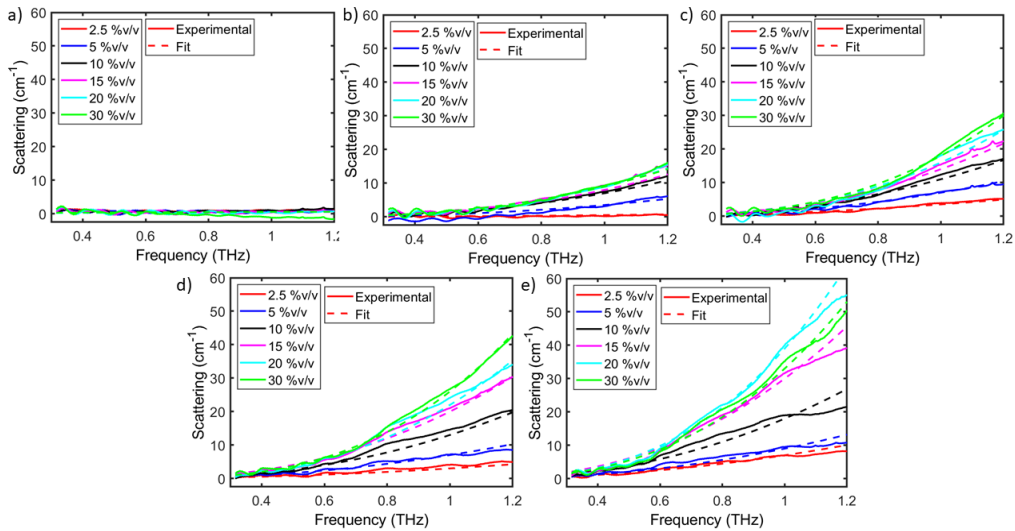

Figure 4: Power law fitted to scattering spectra a) 38 – 45 µm, b) 90 – 106 µm, c) 125 – 150 µm, d) 150 – 180 µm & e) 180 – 212 µm.

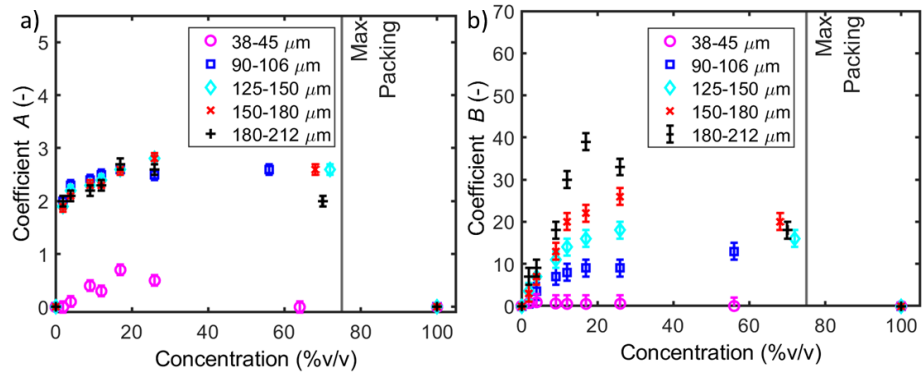

Figure 5: Coefficients a) *A* & b) *B* from applying the power law (Eq. 2) to the estimated scattering contributions. Values plotted are the mean and estimated confidence intervals. Additionally Maximum packing of spheres (face-centered cubic lattice, 74%v/v) was plotted as a vertical line [37].

We also investigated the effect of scattering on the behavior of the effective refractive index. The frequency-averaged refractive index for all concentration levels of borosilicate microspheres can be seen in Figure 6. According to the linear mixing model, the refractive index is

expected to be linearly dependent on the concentration of microspheres:

$$n_{\text{eff}} = C\, n_{\text{glass}} + (1 - C)\, n_{\text{PTFE}} \qquad (9)$$

varying between the refractive index of PTFE ($n_{\text{PTFE}} = 1.45$) and solid borosilicate glass ($n_{\text{glass}} = 2.45$). At low concentrations, the refractive index indeed increases linearly. However, at higher concentrations > 15 %v/v, it becomes nonlinear. At high concentrations, > 50 %v/v, close to the maximum packing density, the refractive index exceeds the linear model.

This behavior may be explained by considering the path length of scattered photons. At low concentrations photons are less likely to undergo multiple scattering, so that scattering events increase linearly with concentration. At the highest concentrations, photons experience multiple scattering, leading to slower propagation through the sample and thus a higher refractive index.

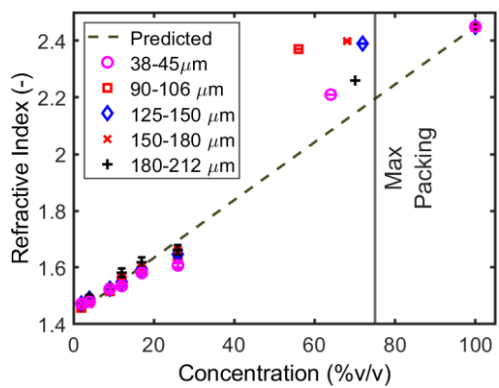

Figure 6: Frequency averaged refractive index for all particle sizes and concentrations. The values plotted are the mean and standard deviation ($n = 10$, < 30%v/v). Additionally Maximum packing of spheres (face-centered cubic lattice, 74%v/v) was plotted as a vertical line [37].

## Conclusion

In this study, we conducted a systematic investigation of THz scattering in granular compacts with varying particle sizes and concentrations. We fabricated granular compacts formed of spherical particles in a solid medium with negligible porosity. Particle size and concentration was controlled and varied with high accuracy and reproducibility. THz-TDS was used to investigate the effect of varying particle size and concentration on scattering. Using the measured optical properties of borosilicate glass, scattering contributions to the loss coefficient were isolated, allowing the effect of particle size and concentration on scattering to be quantified.

It was found that scattering increased linearly at lower concentrations. Scattering saturated at higher concentrations, reaching a maximum level dependent on the particle size. However, normalized scattering curves showed that the onset of saturation is independent of the particle size.

By applying the power law function to the scattering component, we reveal the dependence of the power law coefficients on particle size and concentration. It was found that the pre-exponential factor increased as a function of both concentration and particle size. The dependence on concentration was linear at low concentrations until the onset of saturation. The exponent was similarly dependent on concentration but independent of particle size.

The effective refractive index becomes sublinear with increasing particle concentration. At concentrations close to maximum density, it exceeds the linear model. This finding indicates that calculations based on the measured effective refractive index, such as porosity, are liable to have errors in samples with large scattering objects and high concentrations.

These findings may be explained by considering ballistic versus scattered photon propagation through the compact. At lower particle concentrations, most photons travel ballistically (without scattering). As the particle concentration increases, scattering increases linearly until the onset of saturation due to multiple scattering. As the concentration of microspheres increases further and approaches fully dense packing, scattering falls again as photons increasingly travel ballistically through the glass.

# Appendix

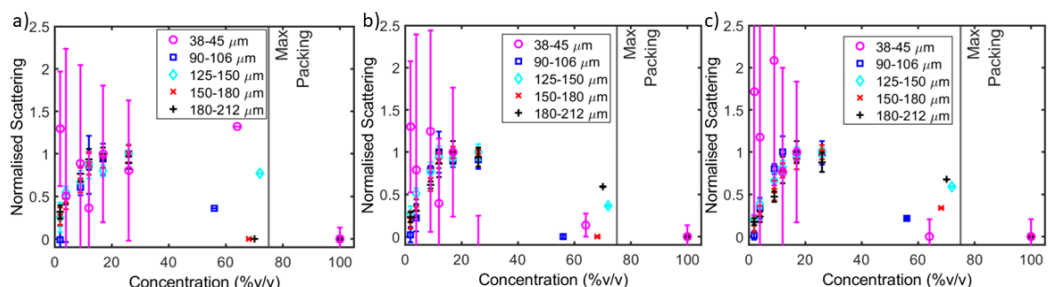

Figure 3: Normalized scattering contributions for a) 0.6 THz, b) 0.8THz & c) 1 THz. Values plotted are the mean with error bars representing the standard deviation ($n = 10$).

# Acknowledgments

The work at NPL was supported by the UK government's Department for Science, Innovation and Technology (DSIT).

The authors would like to thank and acknowledge the contribution of Dr Andrew Burnett for valuable discussions.